\documentclass[12pt]{iopart}
\usepackage{graphicx}	
\usepackage{subfigure}
\usepackage{dcolumn}  					
\usepackage{bm}
\usepackage{iopams}

\usepackage{hyperref}
\hypersetup{colorlinks=true, citecolor=blue, urlcolor=blue, linkcolor=blue}

\usepackage{xcolor}    
\usepackage{soul}      

\usepackage{placeins}

\begin{document}

\title{Transportable strontium lattice clock with $4 \times 10^{-19}$ blackbody radiation shift uncertainty}

\author{I.~Nosske, C.~Vishwakarma, T.~L\"ucke, J.~Rahm, N.~Poudel, S.~Weyers, E.~Benkler, S.~D\"orscher, C.~Lisdat}

\address{Physikalisch-Technische Bundesanstalt, Bundesallee 100, 38116 Braunschweig, Germany}
\eads{\mailto{ingo.nosske@ptb.de}, \mailto{christian.lisdat@ptb.de}}
\markboth{Transportable strontium lattice clock with $4 \times 10^{-19}$ BBR shift uncertainty}{}

\begin{abstract}
We describe a transportable optical lattice clock based on the \mbox{$^1\mathrm{S}_0 \rightarrow {^3\mathrm{P}_0}$} transition of lattice-trapped $^{87}$Sr atoms with a total systematic uncertainty of \mbox{$2.1 \times 10^{-18}$}. 
The blackbody radiation shift, which is the leading systematic effect in many strontium lattice clocks, is controlled at the level of $4.0 \times 10^{-19}$, as the atoms are interrogated inside a well-characterised, cold thermal shield.
Using a transportable clock laser, the clock reaches a frequency instability of about $5 \times 10^{-16}/\sqrt{\tau/\mathrm{s}}$, which enables fast reevaluations of systematic effects. 
By comparing this clock to the primary caesium fountain clocks CSF1 and CSF2 at Physikalisch-Technische Bundesanstalt, we measure the clock transition frequency with a fractional uncertainty of $1.9\times 10^{-16}$, in agreement with previous results. 
The clock was successfully transported and operated at different locations.
It holds the potential to be used for geodetic measurements with centimetre-level or better height resolution and for accurate inter-institute frequency comparisons.
\end{abstract}
\maketitle

\section{Introduction}

State-of-the-art optical atomic clocks have reached fractional systematic uncertainties of few $10^{-18}$ and below \cite{hun16, mcg18, bre19, hua22, tof24, aep24, arn24, hau25}. This progress in the field of optical frequency metrology has triggered active discussions of the redefinition of the SI second \cite{dim24}.
One of the mandatory criteria for such a redefinition is the validation of optical clocks by inter-institute comparisons \cite{dim24, icon24}.
A transportable optical clock can be used as a frequency reference for achieving this goal.
Another field that benefits from the development of highly accurate transportable clocks is chronometric geodesy \cite{meh18}, which can rival the uncertainties of geopotentials determined by state-of-the-art geodetic techniques, typically equivalent to a few centimetres of physical height \cite{den17}.
This will facilitate improved and more consistent height reference systems and allow for better Earth monitoring.
These and other applications, however, require transportable clocks with performance comparable to the best laboratory-based systems.
Several groups are developing transportable optical clocks \cite{kol17, ori18, hua20, ohm21, stu21, guo21, kal22, liu23, zen23, bra24}, but few have reached similar performance as the best laboratory-based clocks \cite{ohm21, bra24}.

Here, we present our second-generation transportable strontium optical lattice clock Sr4
that was already used in off-site measurements \cite{icon24}.
Several fundamental design changes improved the performance compared to its predecessor Sr2 \cite{kol17}: 
The uncertainty of the blackbody radiation (BBR) shift, often representing the largest uncertainty contribution in optical lattice clocks \cite{mcg18, aep24, hob20a, got23, li24}, is reduced to $4.0 \times 10^{-19}$ by transporting the atoms into a cooled copper shield for interrogation, inspired by a design from RIKEN \cite{ush15}. 
This represents an improvement of a factor of about 2 compared to the lowest BBR shift uncertainties reached in strontium lattice clocks \cite{aep24, ush15}, and more than an order of magnitude improvement compared to our previous transportable clock \cite{kol17}.
Unlike the approach in \cite{has25}, it does not require moving mechanical parts.
A new transportable clock laser \cite{her22} allows for a clock instability of about $5 \times 10^{-16}/\sqrt{\tau/\mathrm{s}}$, which reduces the averaging time $\tau$ required for reevaluations of systematic effects by more than one order of magnitude compared to our previous configuration \cite{kol17, hae20}. 
The estimated total systematic uncertainty of the clock is $2.1 \times 10^{-18}$, which is in the range of state-of-the-art laboratory systems and represents a significant step towards enabling the applications outlined above.
We determine the $^{87}$Sr clock transition's absolute frequency by comparison to the primary caesium fountain clocks CSF1 and CSF2 at Physikalisch-Technische Bundesanstalt (PTB) \cite{wey18}.

\section{Clock setup}\label{sec:ClockSetup}

{\begin{figure*}[tb]
\centering
\includegraphics[width=1\linewidth]{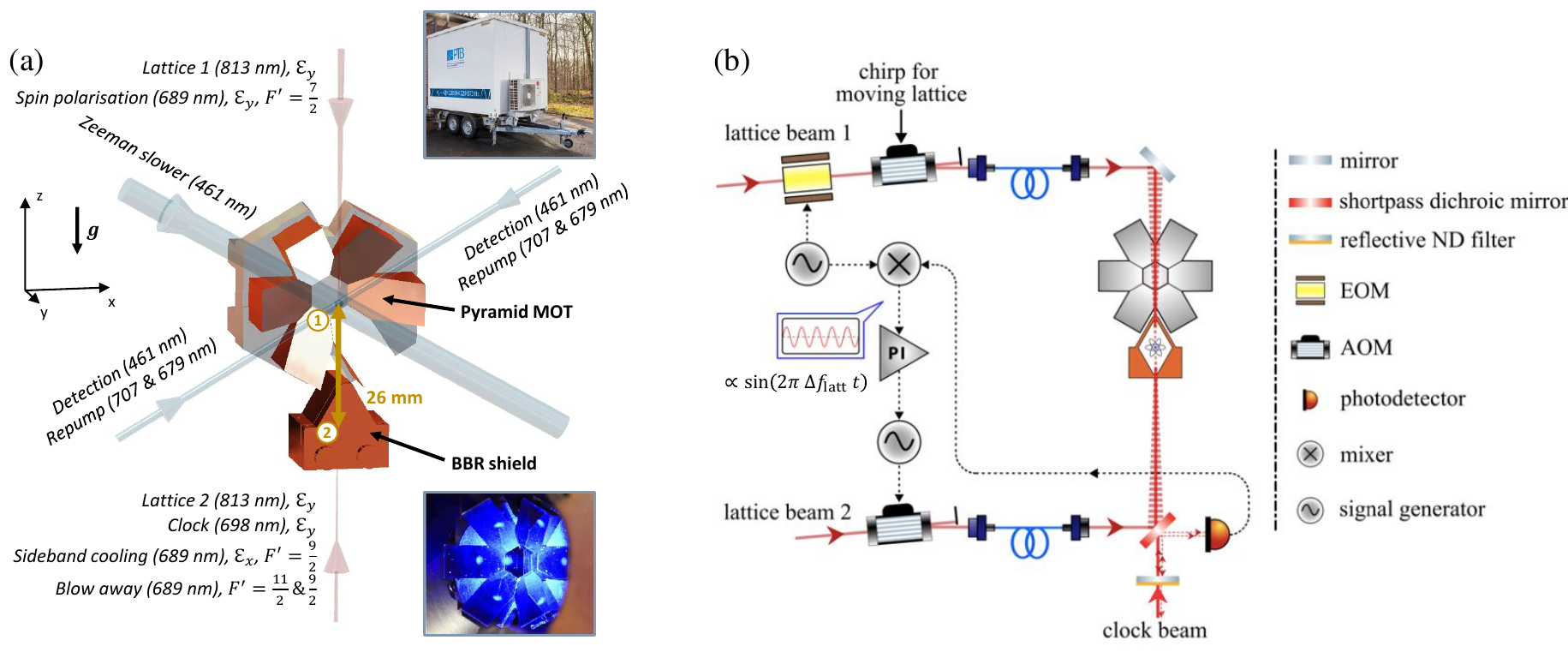}
\caption{Layout of the core components of the physics package. (a) A model of the pyramid MOT (1) and the BBR shield (2) in the centre of the main vacuum chamber, together with all laser beams except the MOT cooling beam. 
For the vertical beams the linear polarisation direction is indicated, where this is relevant.
The upper state hyperfine levels of $^{87}$Sr addressed by the \mbox{689 nm} beams are also given. 
The upper inset shows the air-conditioned car trailer of the clock. 
In the lower inset, the central fluorescence of strontium atoms as well as reflections from the mirrors during the first-stage MOT can be seen. 
(b) Optics setup which subsequently enables a moving lattice and Doppler cancellation of the lattice and clock beams, similar to \cite{ush15}. 
The relative phase of the two lattice beams is stabilised while the atoms are in the BBR shield.
The reflective neutral density (ND) filter serves as end reference surface for the path length stabilisation of the \mbox{698 nm} clock light and thus ensures that no Doppler shift between the lattice-trapped atoms and the clock laser beam occurs. For details see text.}
\label{fig:setup}
\end{figure*}}

The physics package for laser cooling and trapping the strontium atoms, the required lasers and the electronic drivers are mounted in an air-conditioned car trailer (see \cite{kol17} for details). 
The trailer also houses a frequency comb that allows to measure the frequency ratio between the fundamental wavelength of the clock laser at \mbox{1397 nm} \cite{her22} and light from the long-distance interferometric fibre links in Europe at \mbox{1542 nm} \cite{sch22a, cli20a}. 
The ratio is measured in a single branch of the comb without uncompensated optical paths \cite{ben19}.
The comb has an additional branch at \mbox{813 nm} to measure the lattice laser wavelength. 
The mass of all components inside the trailer is about 700~kg. 

The ultra-high vacuum system for trapping and interrogation of $^{87}$Sr atoms is pumped by non-evaporable getter and ion getter pumps.
The latter can be switched off for more than one week without a lasting vacuum degradation. 
A strontium beam is generated in an oven by heating metal to about \mbox{500 $^{\circ}$C}.
A level scheme with the relevant electronic transitions in $^{87}$Sr is shown in \ref{app:LevelScheme}.
The atoms are decelerated in a permanent magnet transverse field Zeeman slower operating on the $^1\mathrm{S}_0 \rightarrow {^1\mathrm{P}_1}$ (\mbox{461 nm}) transition \cite{hil14}.
The atoms are trapped \mbox{33 cm} downstream from the oven nozzle in a single 5 cm-diameter ($1/e^2$) beam on the $^1\mathrm{S}_0 \rightarrow {^1\mathrm{P}_1}$ transition in a pyramid MOT \cite{bow19a}, which consists of six silver-coated mirrors made from oxygen-free copper and a central sharp-edged CaF$_2$ prism mounted in the vacuum chamber, see figure~\ref{fig:setup}~(a).
The copper substrates are screwed to an aluminium base plate; the CaF$_2$ prism is held by glue and secured by clamp.
Optical access for additional laser beams, as well as the atomic beam, is granted by cusps between the six mirror elements.

A water-cooled anti-Helmholtz coil pair generates the MOT magnetic field gradients.
Three orthogonal Helmholtz coil pairs provide precise magnetic field compensation and control.  
The centre of the physics package is enclosed by a single-layer \mbox{1.5 mm}-thick $\mu$-metal shield.
Despite having holes for the required laser beams, cables and vacuum system connections, the $\mu$-metal shield provides a magnetic shielding factor at its centre of $\gtrsim$100, which enables re-trapping of atoms in the optical lattice without the need for new field compensation after transporting the clock to another site. 

During a second-stage laser cooling of $^{87}$Sr atoms to a few $\mu$K via the $^1\mathrm{S}_0 \rightarrow {^3\mathrm{P}_1}$ (\mbox{689 nm}) transition \cite{muk03},
the atoms are loaded into an optical lattice formed by two counter-propagating beams at \mbox{813 nm}.
The lattice is operated close to the $E$1 magic frequency for the clock transition. 
Below the pyramid MOT assembly, a BBR shield \cite{ush15} is installed
through which the lattice is transmitted via two holes with diameters of about \mbox{1 mm}.
Atoms are moved downwards into the centre of the \mbox{20 mm}-long copper shield by a frequency chirp on one of the lattice beams.
The shield is attached to a pulse tube refrigerator and can be cooled to below \mbox{100 K}.
It not only provides a well-controlled thermal environment with reduced BBR, but also shields the atoms from potentially existing stray electric fields, and from the hot atomic beam and BBR from the oven. 

The clock laser with its reference resonator \cite{her22} is located outside the trailer in a protected place to avoid vibrations from the air conditioning system and water chillers in the trailer.
Its ultra-stable light at \mbox{1397 nm} is delivered to the physics package via fibre with noise cancellation. There, it is frequency converted in a fibre-to-free-space periodically-poled lithium niobate (PPLN) waveguide frequency doubler to the interrogation wavelength of \mbox{698 nm}.
The endpoint of the fibre noise cancellation at \mbox{1397 nm} thereby serves as reference point for an optical path length stabilisation of the switched light distribution to the atoms at \mbox{698 nm} (see figure~S4~b) in \cite{her22}).

The clock laser beam to interrogate the $^1\mathrm{S}_0 \rightarrow {^3\mathrm{P}_0}$ transition is overlapped with the lattice and focussed (like the lattice beams) at the BBR shield centre.
The $1/e^2$ beam waists are \mbox{110 $\mu$m} and \mbox{66 $\mu$m}, respectively. 
Furthermore, laser beams at \mbox{689 nm} are superimposed with the lattice for axial sideband cooling, spin polarisation and removal of atoms in unwanted states.
Where relevant, the polarisations of the beams are indicated in figure~\ref{fig:setup}~(a).
A magnetic field is applied with varying strength along the $y$ direction during the following steps.

Atoms are moved into the centre of the BBR shield in a \mbox{60 ms}-long transfer phase. The frequency of lattice beam~1 is detuned by at maximum \mbox{1.6 MHz} with respect to lattice beam~2 to form a moving lattice that shifts the atoms by \mbox{26 mm}.
During the preparation and spectroscopy stages inside the BBR shield, the relative phase of the two lattice beams is stabilised by an acousto-optic modulator (AOM).
An electro-optic modulator (EOM) imprints sidebands to generate a Pound-Drever-Hall like error signal, see figure~\ref{fig:setup}~(b) for details.
In the BBR shield, the atoms are axially sideband cooled and pumped to the $m_F = \pm9/2$ stretched hyperfine ground-state. 
To prepare a pure spin sample, atoms in either of the two stretched $m_F$ states are transferred to the $^3\mathrm{P}_0$ state by a 60 ms long $\pi$ pulse on the clock transition. This is done in a bias magnetic field of about \mbox{0.41 mT} that splits the $m_F = \pm 9/2$, $\Delta m_F = 0$ clock transitions by \mbox{4.0 kHz}. 
Remaining ground-state atoms are blown away by an intense beam on the 689 nm transition.
After adiabatically reducing the lattice depth to the desired value (typically $15...50 \, E_\mathrm{rec}$ with $E_\mathrm{rec} = h \times 3.47~\mathrm{kHz}$ the recoil energy of a lattice beam photon), we perform Rabi spectroscopy by a typically \mbox{500 ms} long $\pi$ pulse on the clock transition at a Zeeman splitting of about \mbox{540 Hz}.

After the clock interrogation, the atoms are moved back up to the MOT position.
Here, the atoms in the $^1\mathrm{S}_0$ state are removed from the trap by radiation pressure. 
The arising \mbox{461 nm} fluorescence is collected by a photomultiplier tube via a $3^{\prime\prime}$ diameter off-axis parabolic mirror, which has a central \mbox{35 mm} diameter hole along the focal axis to give optical access for the MOT cooling beams.
Following the repumping of the remaining atoms from the $^3\mathrm{P}_{0}$ to the $^1\mathrm{S}_{0}$ state, the same detection procedure is repeated. The signal background is recorded in a third detection.
The excitation probability is determined from the three signals. 
Frequency scans over the clock transition are shown in figure~\ref{fig:ClockSpectro}~(a).

{\begin{figure*}[tb]
\centering
\includegraphics[width=1\linewidth]{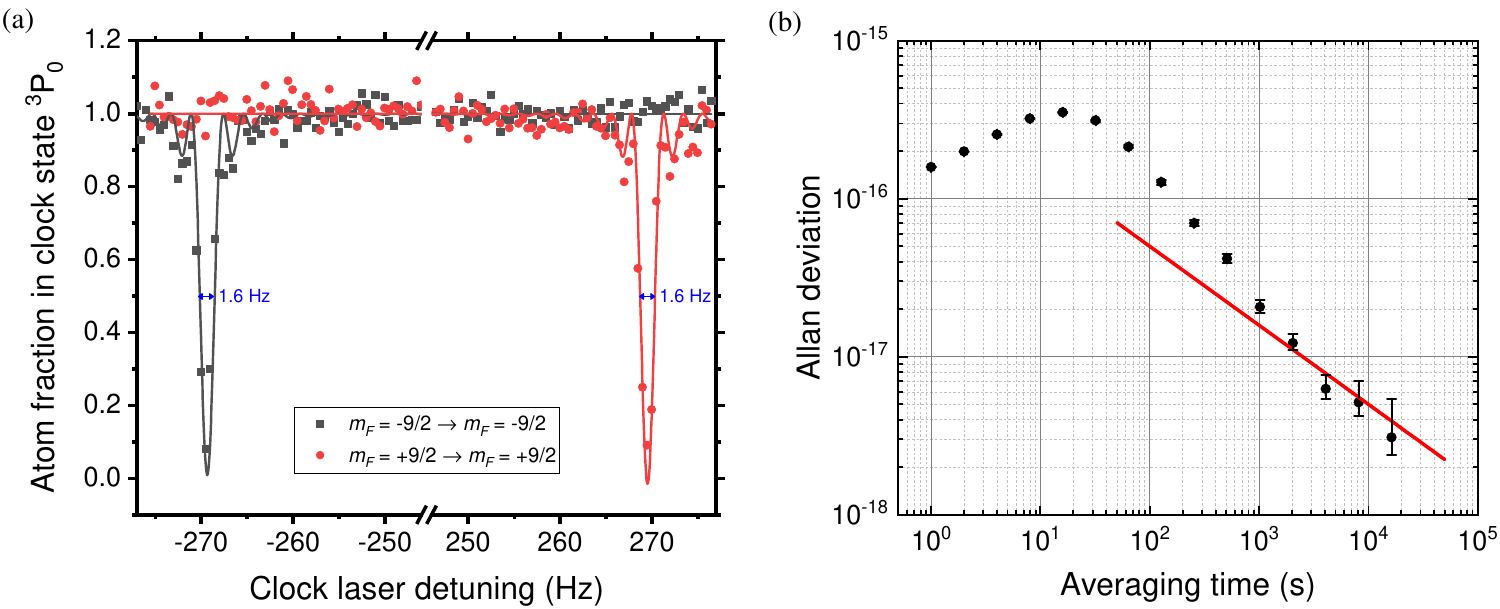}
\caption{Spectroscopy of and stabilisation on the \mbox{698 nm} clock transition. (a) 
Scans of the clock transition of atomic samples spin-polarised in the $m_F = \pm 9/2$ states with \mbox{500 ms} Rabi interrogation pulses.
The solid lines are fits with the expected line shape.
(b) Frequency stability of a comparison between the transportable lattice clock Sr4 and the laboratory clock Sr3 \cite{sch22b}, which is much more stable thanks to an ultra-stable silicon resonator \cite{mat17a}. The line indicates the Sr4 instability of $5 \times 10^{-16}/\sqrt{\tau /\mathrm{s}}$. For this measurement in Sr4 a dead time of 1192~ms and a Rabi time of 450~ms were used.}
\label{fig:ClockSpectro}
\end{figure*}}

For stabilising the clock laser frequency to the atomic resonance, we subsequently interrogate the $m_F = \pm 9/2$ transitions at the low and high frequency half width points.
From the four observed excitation probabilities, we estimate the frequency offset of the clock laser from the atomic resonance and the line splitting induced by the magnetic field.
By observing the temporal development of the frequency corrections applied to the clock laser, we also estimate the drift rate of the clock laser resonator.
To minimise lock errors, we apply a feed forward compensation of the cavity drift \cite{fal11}.
 
Using the transportable clock laser \cite{her22}, the clock achieves an instability of $5 \times 10^{-16}/\sqrt{\tau/\mathrm{s}}$, see figure~\ref{fig:ClockSpectro}~(b). 
For evaluations of systematic frequency shifts in the clock, we interleave two clock stabilisations with different clock parameters, e.g. different lattice trap depths or atom numbers \cite{alm15}.
Here, we observe an asymptotic instability of about $7 \times 10^{-16}/\sqrt{\tau/\mathrm{s}}$, enabling fast characterisations of systematic effects with small statistical measurement uncertainties.

\section{BBR shield and shift}\label{sec:BBRShieldFrequencyMapping}

The BBR shift often causes the largest uncertainty contribution in state-of-the-art optical lattice clocks \cite{mcg18, hob20a, got23, aep24, li24} and therefore requires special attention. 
Critical parameters are the representative temperature $T$ of the environment and the atomic response to the BBR field.
In the electric dipole ($E$1) approximation, the BBR shift $\Delta \nu_\mathrm{BBR}(T)$ is often described by the 
sum of the so-called static contribution $\Delta \nu^\mathrm{(stat)}(T) \propto T^4$ \cite{mid12a} and the dynamic contribution $\Delta \nu^\mathrm{(dyn)}(T)$. Here, we express the latter as $\Delta \nu^\mathrm{(dyn)}(T) = \Delta \nu^\mathrm{(dyn)}(T_0) \left( T/T_0 \right)^6 f(T/T_0)$ with
\begin{equation}
f(T/T_0) = \frac{\eta_6 + \eta_8 \left( T/T_0 \right)^2 + \eta_{10} \left( T/T_0 \right)^{4}}{\eta_6 + \eta_8 + \eta_{10}}
\label{eq:dyn}
\end{equation}
to reproduce the dynamic contribution at $T = T_0 \equiv 300 \, \mathrm{K}$ and facilitate computing the uncertainty.
We update the coefficients $\eta_i$ from \cite{lis21a} by scaling them with the fractional difference with respect to the recently reevaluated value $\Delta \nu^\mathrm{(dyn)}(300 \, \mathrm{K}) = -153.06(33)~\mathrm{mHz}$ \cite{aep24}, where the number in parentheses is the uncertainty referred to the corresponding last digits of the result. The such determined coefficients are $\eta_6 = -0.132\,16~\mathrm{Hz}$, $\eta_8 = -0.012\,31~\mathrm{Hz}$ and $\eta_{10} = -0.008\,58~\mathrm{Hz}$ and  agree with the full calculation in \cite{aep24} within $1 \times 10^{-19}$ for $T \leq 300~\mathrm{K}$ \cite{kim25}.
The $M$1 BBR shift only amounts to $5.6 \times 10^{-20}$ at $300 \, \mathrm{K}$ \cite{por06, tan24} and is neglected in the following analysis. 

{\begin{figure*}[tb]
\centering
\includegraphics[width=0.5\linewidth]{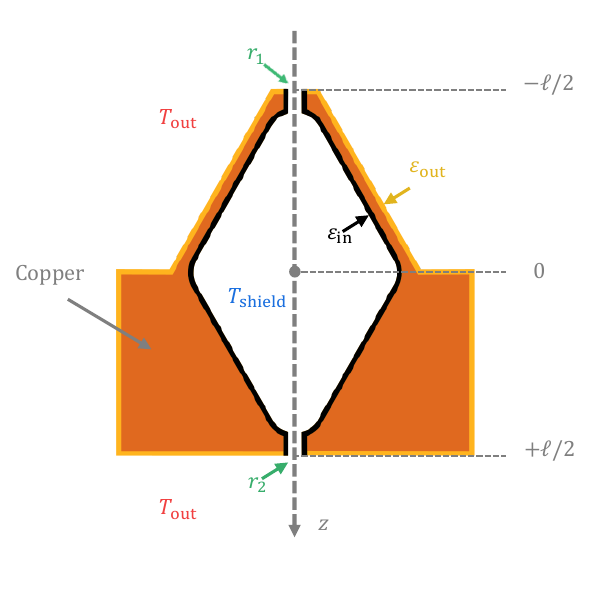}
\caption{Cut through the BBR shield in the $xz$ plane with indications of the hole positions, hole radii, temperatures and emissivities that are relevant for the BBR shift determination.}
\label{fig:shieldgeometry}
\end{figure*}}

During interrogation, the atoms are mainly exposed to the BBR from the shield at temperature $T_\mathrm{shield}$.
However, BBR from the outside $T_\mathrm{out}$ enters through the holes in the shield and may illuminate the atoms directly or after scattering from the walls.
The solid angle $\Omega$ under which the atoms see the exterior determines the amount of direct line-of-sight room temperature BBR the atoms are exposed to. 
It depends on the distance between atoms and holes, $z \pm \frac{\ell}{2}$, that is given by the atomic position $z$ with respect to the centre of the shield at $z=0$ and the length of the BBR shield $\ell$ (figure~\ref{fig:shieldgeometry}).
For hole radii $r_i$, the fractional solid angle is given by:
\begin{equation}\label{eq:SolidAngle}
\fl \frac{\Omega(z)}{4\pi} = \frac{1}{2} \left[ 1 - \sin\left( \arctan \left( \frac{z + \frac{\ell}{2}}{r_1} \right) \right) \right] + \frac{1}{2} \left[ 1 + \sin\left( \arctan \left( \frac{z - \frac{\ell}{2}}{r_2} \right) \right) \right]
\end{equation}
\noindent To determine the contribution of the room temperature BBR, the radii of both holes were accurately measured to be \mbox{0.484(6) mm} (\ref{app:SolidAngle}), resulting in a solid angle of $4 \pi \cdot 1.17(3) \times 10^{-3}$ at the BBR shield centre. 

The inside of the BBR shield is coated by a high-emissivity coating. Based on the hemispherical reflectance in the \mbox{2.6 ... 17 $\mu$m} range certified by the supplier (Ultra Black\textsuperscript{TM} by Acktar), we find an emissivity of the inner coating $\epsilon_\mathrm{in} = 0.926(43)$.
Hence, also BBR from the outside that is scattered on the inner walls may interact with the atoms.
This increases the effective solid angle under which the atoms see the holes to \cite{mid11,abd19}
\begin{equation}
\frac{\Omega_\mathrm{eff}(z)}{4 \pi} = \frac{1}{1 + \left( \frac{4 \pi}{\Omega(z)} -1 \right) \epsilon_\mathrm{in}}~.
\end{equation}
\noindent
The position-dependent BBR shift in the shield $\Delta \nu_\mathrm{BBR}^\mathrm{shield}$ is expected to be:
\begin{equation}\label{eq:BBRexpectation}
\Delta \nu_\mathrm{BBR}^\mathrm{shield}(z) = \Delta \nu_\mathrm{BBR}(T_\mathrm{shield})\left( 1 - \frac{\Omega_\mathrm{eff}(z)}{4\pi} \right) + \frac{\Omega_\mathrm{eff}(z)}{4\pi}  \Delta \nu_\mathrm{BBR}(T_\mathrm{out})
\end{equation}
In order to calibrate the position of the atoms in the BBR shield and verify the model in (\ref{eq:BBRexpectation}), we interleave clock stabilisations with the atoms residing either at \mbox{$z = 0 \, \mathrm{mm}$} or at another distance $z$ in or around the BBR shield.
The measured fractional frequency differences are plotted in figure~\ref{fig:InterleavedResults} for $T_\mathrm{shield} = -50.1(1) \, ^{\circ}\mathrm{C}$ and $T_\mathrm{out} = 21(1) \, ^{\circ}\mathrm{C}$.
The error bars are given by the statistics of the measurements.
Since the bias magnetic field varies over the probed region, we have corrected the measured shifts for the difference of the second-order Zeeman shifts (section~\ref{sec:SecondOrderZeeman}), which amount to at most $1.7 \times 10^{-17}$.
In order to check for differential lattice light shifts, we compared the clock frequency at two lattice intensities ($I_2/I_1 = 1.6$) at the centre and inside the upper hole of the BBR shield. As they are measured to be zero-compatible within their statistical measurement uncertainties of $< 2 \times 10^{-17}$, they are not included in this analysis.
Other effects have been neglected.

{\begin{figure*}[tb]
\centering
\includegraphics[width=1\linewidth]{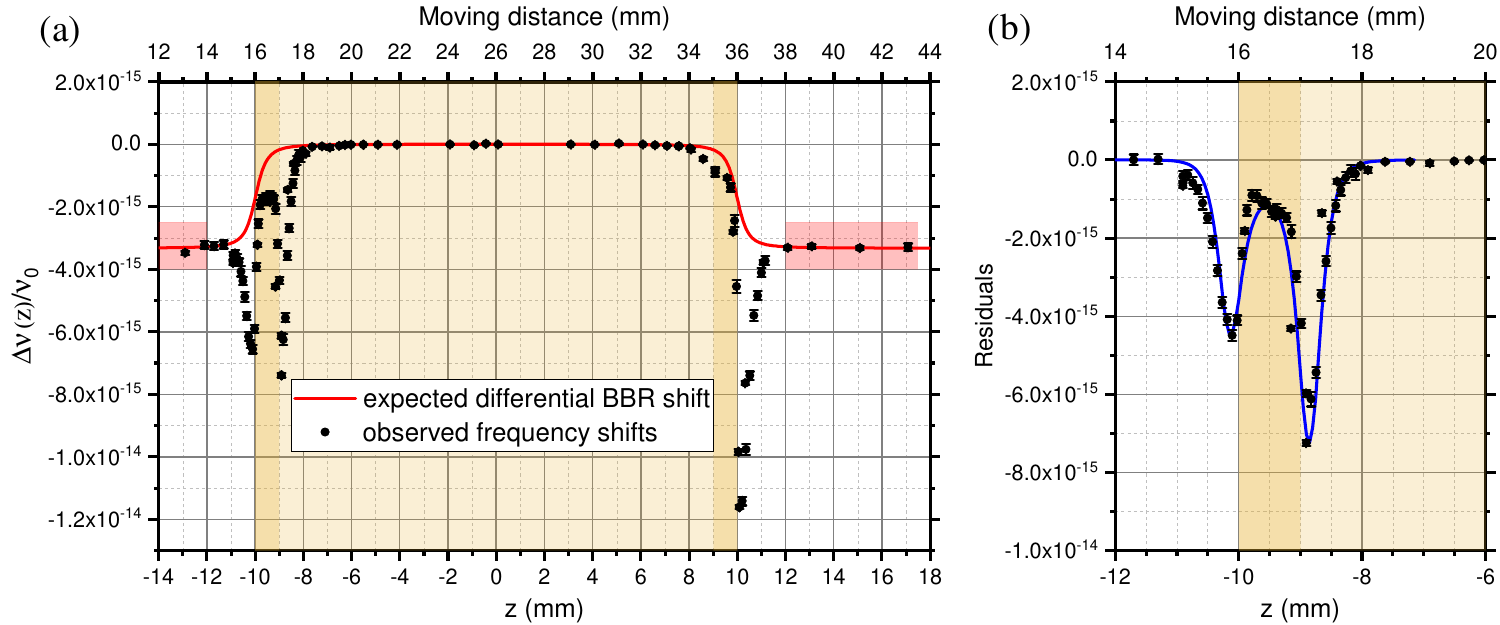}
\caption{Differential frequency shifts in and around the BBR shield. (a) Position dependent fractional frequency shift relative to the centre of the BBR shield 
 at $T_\mathrm{shield} = -50 \, ^{\circ}\mathrm{C}$ and $T_\mathrm{out} = 21 \, ^{\circ}\mathrm{C}$ (dots). 
  The red curve is the expected differential BBR shift according to (\ref{eq:BBRexpectation}).
  The light-yellow region denotes the BBR shield interior, the dark-yellow regions its two \mbox{1 mm}-deep holes.
  The red regions outside the BBR shield indicate the data points that were used to calculate the experimentally observed differential BBR shift, see text.
  (b) Residuals between expected BBR and observed frequency shifts (dots). The blue line is a simulation of Stark shifts due to surface potentials.
  }
\label{fig:InterleavedResults}
\end{figure*}}

The difference of frequency shifts for atoms inside and outside the BBR shield matches the expectations 
for the BBR shift from (\ref{eq:BBRexpectation}).
The weighted average of the data points for atoms outside the BBR shield and at least \mbox{2 mm} away from the nearest outer hole edge (see the red-shaded regions in figure~\ref{fig:InterleavedResults} (a)) is $-3.33(3) \times 10^{-15}$, which agrees with the expected differential BBR shift $\Delta \nu_\mathrm{BBR}^\mathrm{shield}\left(|z| > \ell/2 + 2 \, \mathrm{mm}\right) - \Delta \nu_\mathrm{BBR}^\mathrm{shield}(0) = -3.32(7) \times 10^{-15}$. 
The uncertainty of the latter value results from the uncertainty of $T_\mathrm{out}$.
We neglect the influence of BBR from the outer shield surfaces, as they are polished and have a small emissivity $\epsilon_\mathrm{out} \approx 0.03$.

In contrast to the expected smooth development of the shift from the inside to the outside of the BBR shield, we observe localised peaks at the inner and outer edges of the holes.
As the cause for the peaks at the outer edges, we consider surface potentials due to the transition from coating to polished copper on the outside of the shield.
For a sufficiently thick coating, no further shifts due to differences in the work function on the inner surface would be expected.
This behaviour is mostly observed for the hole at $z >0$ and can be well reproduced by a finite element method (FEM) simulation that incorporates a small varying surface potential on the inner edge of the bore caused by a reduced coating thickness.
However, for $z<0$ much stronger shifts are observed on the inside of the hole.
These can be modelled if a linear variation of the surface potential along the bore and an offset from the hole axis are assumed, see figure~\ref{fig:InterleavedResults}~(b).
We suspect that the coating in this bore has not reached its intended thickness as in the other, easier to cover regions of the BBR shield, and causes these patch potentials.

For distances to the hole edge $\Delta z$ large compared to the bore radius $r$, the shift rapidly falls off with an approximate $\propto \Delta z^{-4}$ scaling. 
According to fits of the data, the combined residual shift from the two inner hole edges at the BBR shield centre is up to about $-1 \times 10^{-19}$.

We expect temperature inhomogeneities of the BBR shield due to residual absorption of room temperature BBR and absorption of lattice light near the bores. 
The latter is apparent due to an increase of the shield temperature by \mbox{100 mK} when the lattice is turned on and is attributed to suboptimal alignment.
In order to assess these effects, we perform a FEM simulation of the actual geometry of the copper BBR shield, including a slightly reduced thermal conductivity between the main body and the top lid, which is cold-welded to the bottom with a \mbox{100 $\mu$m} thick indium foil.
The temperature at the bottom of the BBR shield is fixed. 

{\begin{figure*}[tb]
\centering
\includegraphics[width=0.5\linewidth]{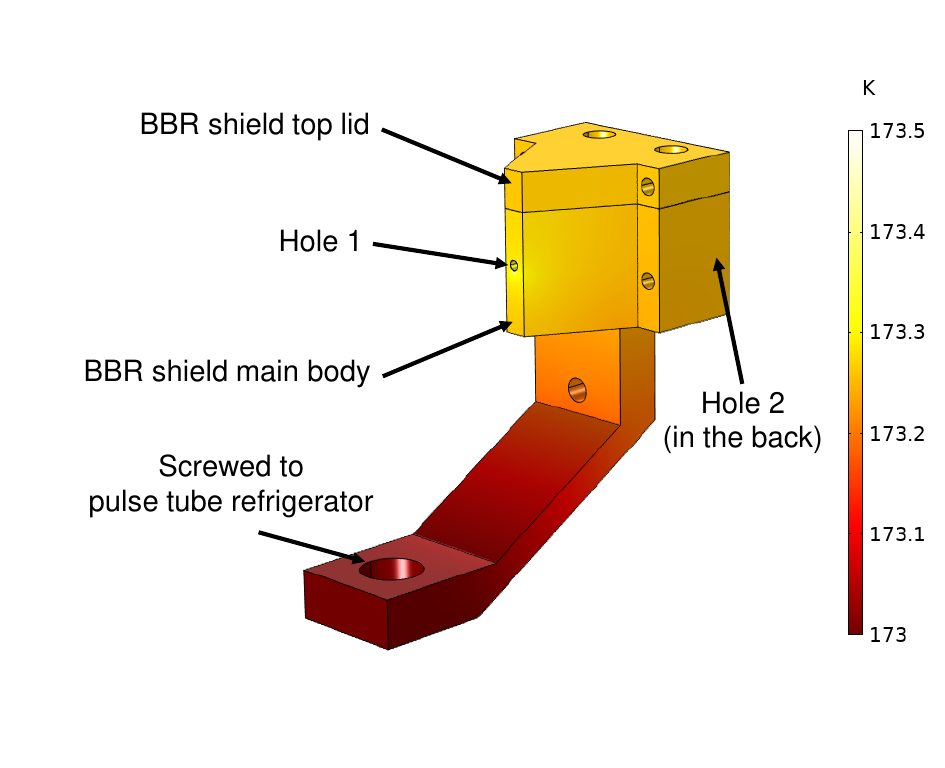}
\caption{FEM simulation of the temperature of the BBR shield.}
\label{fig:BBR_shield_temperature_simulation}
\end{figure*}}

Due to heat intake from the room temperature environment, a cooled BBR shield ($T_\mathrm{shield} < T_\mathrm{out}$) is slightly warmer at its top than at its bottom. 
According to the simulation the maximum temperature difference is $11.4~\mathrm{mK}$ for our geometry.
To account for the heating by the lattice laser, we adjust the absorbed power to match the observed temperature increase of the shield.
The observed heating is consistent with the absorption of about 1\% of the lattice beam power.
The result of the simulation is shown in figure~\ref{fig:BBR_shield_temperature_simulation}.

We see that the two points where part of the lattice beams are partially absorbed exhibit a temperature increase by about \mbox{0.5 K} above the BBR shield temperature measured at the Pt100 positions. 
As this temperature steeply falls off on a length scale of a few \mbox{100 $\mu$m}, it only leads to a negligible atomic BBR shift change of $< 1 \times 10^{-19}$ even at room temperature (and less for lower temperatures), if compared to a homogeneous temperature environment. 
However, the laser heating also leads to smaller temperature gradients on larger length scales in the BBR shield.
According to the simulation, we estimate the difference between the warmest and the coldest point of the BBR shield -- after removing the warm regions around the lattice beam absorption spots -- to be \mbox{53 mK} (between room temperature and at $-100 \, ^{\circ}\mathrm{C}$), which is considerably larger than the thermal inhomogeneity caused by room temperature BBR absorption alone. 
Assuming a rectangular probability distribution of the true temperature value in this range \cite{gum08}, we arrive at a standard uncertainty of \mbox{15 mK}.

In combination with a calibration uncertainty of the Pt100 sensors, which are mounted at the sides of the shield, of \mbox{12.5 mK} and a \mbox{4 mK} uncertainty associated with the measurement bridge, we find an uncertainty of \mbox{20 mK} for $T_\mathrm{shield}$ in the range between room temperature and $-100 \, ^{\circ}\mathrm{C}$.
We correct the measured temperatures for the small difference between the thermodynamic temperature and the ITS-90 temperature scale, $T - T_{90}$ \cite{gai22}, which e.g. for \mbox{223 K} corresponds to $-3.2(2)~\mathrm{mK}$ and for \mbox{173 K} to $-6.6(3)~\mathrm{mK}$.

{\begin{figure*}[tb]
\centering
\includegraphics[width=0.6\linewidth]{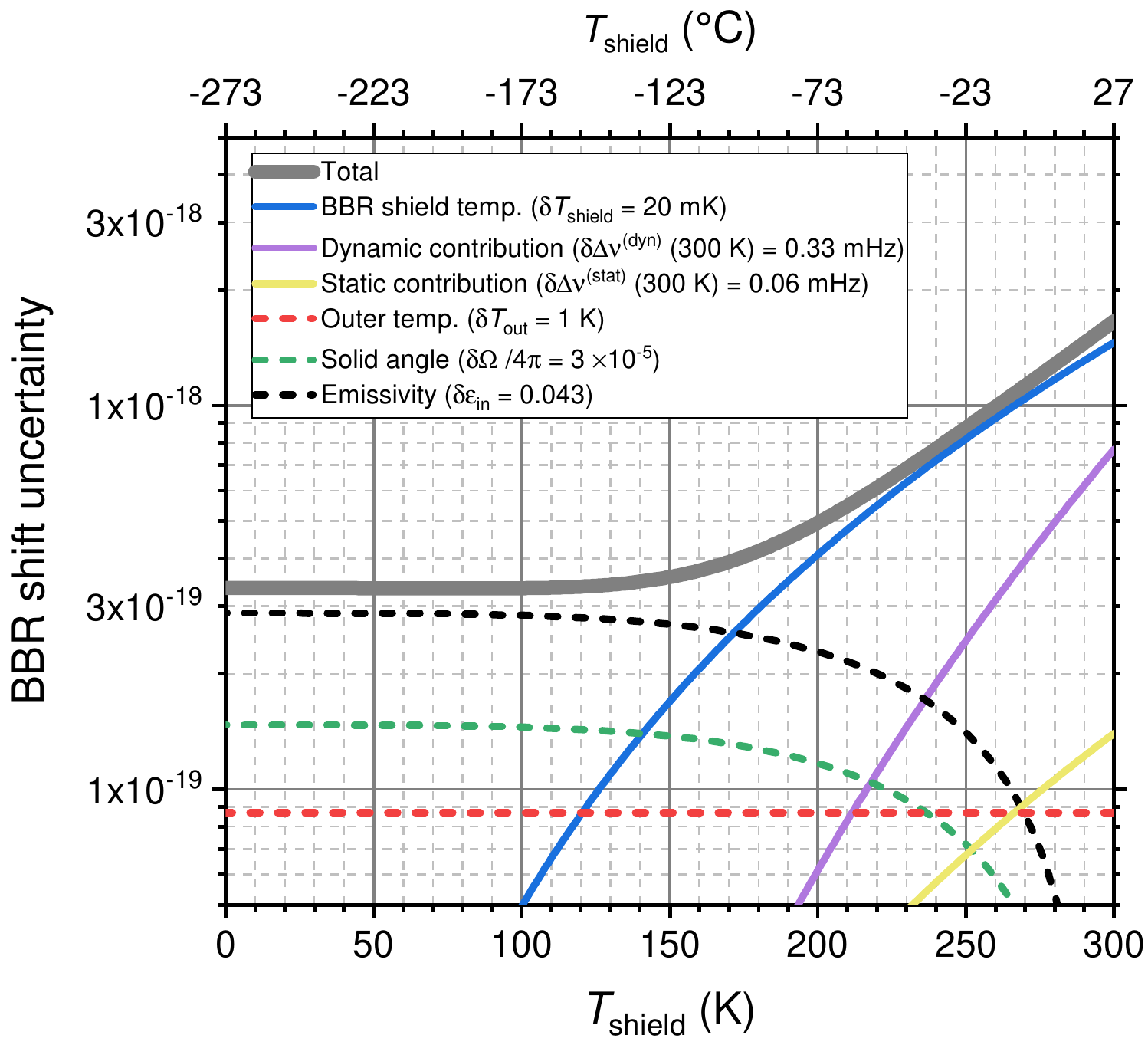}
\caption{BBR shift uncertainty contributions associated with uncertainties of different quantities versus $T_\mathrm{shield}$. The solid (dashed) lines denote uncertainties related to the BBR shift from $T_\mathrm{shield}$ ($T_\mathrm{out}$).}
\label{fig:uncertainty_contributions}
\end{figure*}}

The BBR shift uncertainty contributions associated with uncertainties of the parameters $T_\mathrm{shield}$, $\Delta \nu^\mathrm{(dyn)}$, $\Delta \nu^\mathrm{(stat)}$, $T_\mathrm{out}$, $\frac{\Omega}{4\pi}$ and $\epsilon_\mathrm{in}$ are determined by error propagation. 
In figure~\ref{fig:uncertainty_contributions}, they are plotted versus $T_\mathrm{shield}$. 
Between room temperature and $-100 \, ^{\circ}\mathrm{C}$, the uncertainty associated with $T_\mathrm{shield}$ dominates, while for lower temperatures the uncertainty associated with $\epsilon_\mathrm{in}$ dominates. We operated the clock at $-100 \, ^{\circ}\mathrm{C}$, where the total BBR shift uncertainty is $4 \times 10^{-19}$.
An increase by slow variations of $T_\mathrm{shield}$ is avoided by the application of time-resolved corrections of the BBR shift.
Compared to \cite{ush15}, the BBR shift uncertainty was roughly halved mainly by a better-known solid angle to the outer environment.

We note that a reduction of the total BBR shift uncertainty to the $10^{-20}$ regime appears to be feasible with this design type. This would be achievable by, for example, doubling the BBR shield length to $40 \, \mathrm{mm}$ and for $T_\mathrm{shield} \lesssim 100 \, \mathrm{K}$ -- provided that the uncertainty of the BBR shield temperature does not increase. The corresponding additional moving lattice distance of \mbox{10 mm} would only result in an additional dead time of $\sim$$30 \, \mathrm{ms}$ in our system, only marginally affecting its instability by the Dick effect \cite{dic87}.

\section{Other systematic effects}\label{sec:ClockPerformance}

\subsection{Density shift}

In order to measure the density shift, i.e. interactions between the lattice-trapped fermionic $^{87}$Sr atoms during clock spectroscopy, we perform a self-comparison of the clock at interleaved different atom numbers. 
During this measurement the lattice trap depth is held at $34(3) \, E_\mathrm{rec}$. 
The atom number in one of the two clock operation modes is reduced by shortening the preparation pulse which transfers the atoms in the desired $m_F$ level of the $^3\mathrm{P}_0$ state before the actual clock interrogation. 
In order to evaluate the shift at a different atom number $N$ and lattice depth $U$, we use the density shift scaling $\propto N \, U^{3/4}$ \cite{swa12}. 
At our typical atom number and at the lattice depth $16(2) \, E_\mathrm{rec}$, the resulting density shift is $+0.1(4) \times 10^{-18}$. 
Its uncertainty is dominated by the statistical uncertainty of the underlying measurement and includes a smaller contribution due to atom number variations during clock operation of up to 50\%.

\subsection{Lattice light shift}

The lattice beams are derived from a titanium:sapphire laser and spectrally filtered by a volume Bragg grating filter with a 23 GHz bandwidth.
In order to evaluate the lattice light shift, we perform a clock self-comparison at interleaved lattice depths of $16(2) \, E_\mathrm{rec}$ and $35(4) \, E_\mathrm{rec}$ at a lattice frequency close to the $E$1 magic frequency \cite{ush18}.
Presently, the frequency stability of the lattice laser is provided by a reference cavity.
To ensure better reproducibility after transportation, we will lock the lattice laser to the Sr clock laser via the optical frequency comb.
With the lattice light shift model outlined in \cite{bro17} and the values for the differential $E$1 polarisability and the hyperpolarisability from \cite{ush18}, we calculate a lattice light shift of $+7.4(1.7) \times 10^{-18}$ for the stabilization cycle with the shallow lattice depth. 
For the \mbox{$E$2-$M$1} polarisability coefficient, we use the weighted mean $\tilde{\alpha}^{qm}/h = -1.07(14) \, \mathrm{mHz}$ of the coefficients published in \cite{ush18} and \cite{kim23}.
The population distribution required as input to the light shift model is determined by sideband spectroscopy.
For the shallow [deep] lattice, the fraction of atoms in axial states $n_z \textgreater 0$ is 0.16(5) [0.08(5)], the radial temperature is \mbox{1.0(5) $\mu$K} [\mbox{1.5(5) $\mu$K}].
The small density shift difference between the two stabilization cycles is neglected in this analysis.

\subsection{Second-order Zeeman shift}\label{sec:SecondOrderZeeman}

The second-order Zeeman shift is calculated from the shift coefficient from \cite{bot19} and the splitting $\Delta \nu_{\pm 9/2}$ between the two $m_F$ clock transitions, which is tracked during the clock stabilization.
During measurements of the lattice light shift we have seen that this splitting contains only a negligible contribution from a vector light shift of less than \mbox{10 mHz}.
At our typical Zeeman splitting of about \mbox{540 Hz} and daily variations of less than \mbox{0.3 Hz}, the uncertainty of the second-order Zeeman shift amounts to $2.8 \times 10^{-19}$. 
As this is one order of magnitude below the total clock uncertainty, for simplicity we do not apply time-resolved Zeeman shift corrections.

\subsection{Background gas collisions}

The fractional frequency shift of strontium due to collisions with room temperature H$_2$ molecules, which usually dominate in ultra-high vacuum systems, can be calculated by $\Delta \nu_\mathrm{HC}/\nu_0 = -30(3) \times 10^{-18} \, \mathrm{s} / \tau_\mathrm{trap}$ \cite{alv19}. 
In the BBR shield, we measure $1/e$ lifetimes of lattice-trapped $^1\mathrm{S}_0$ atoms ranging from \mbox{9 s} to \mbox{15 s}, while at the MOT position we measure lifetimes of magnetically trapped $^3\mathrm{P}_2$ atoms of up to \mbox{22 s}. 
Due to possible additional loss mechanisms, the measured lattice trap lifetimes only represent a lower bound for the collision-limited trap lifetime. 
We estimate the background gas collision shift to be in the middle between the H$_2$ collision values associated with the two extreme measured lifetimes (\mbox{9 s} and \mbox{22 s}), and the uncertainty such that both values are covered, that is $-2.3(1.0) \times 10^{-18}$.

\subsection{Other contributions}

The clock laser light shift is given by $\Delta\nu_p/\nu_0 = \chi I$, with the light shift at the clock transition of about $\chi = -26~\mathrm{Hz~cm^2/W}$ \cite{lis21a} and the beam intensity \mbox{$I = \frac{11}{9} \frac{2 \pi^3 \, h \, \nu_0^3 \, \tau_{^3\mathrm{P}_0}}{3 \, c^2 \, \tau_p^2}$} for the Rabi $\pi$ pulse duration $\tau_p$, with the Planck constant $h$, the speed of light $c$, and the upper state lifetime $\tau_{^3\mathrm{P}_0} = 118(3) \, \mathrm{s}$ \cite{mun21}. For $\tau_p = 500 \, \mathrm{ms}$ we find a negligible light shift of $-4.2 \times 10^{-20}$, which is also considered as its uncertainty. Due to the conservative estimate 
the influence of the Lamb-Dicke parameter on the Rabi frequency is neglected, as it only increases the probe light shift by less than $20\%$.
Also the differing atomic lifetime recently measured in \cite{lu24a} only causes a shift difference well within the given uncertainty.

The possible size of a DC Stark shift is estimated from the shifts observed near the hole edges discussed in section~\ref{sec:BBRShieldFrequencyMapping}, i.e. \mbox{$-1(1) \times 10^{-19}$}, where the uncertainty is estimated to be equal to the magnitude. 
No electric fields from patch charges or small coating variations on the inner BBR shield walls are expected to be noticeable as the walls have a minimum distance to the atoms of $5 \, \mathrm{mm}$.
Electric fields from the outside are well shielded by the BBR copper shield.

A temporal change of the clock laser cavity drift leads to a servo error when the servo adjusting the feed forward for drift compensation cannot follow fast enough \cite{fal11}. 
For typical operation conditions of our clock, we arrive at a servo error of $0(2) \times 10^{-19}$. 

Line pulling by transitions starting from $m_F \neq \pm 9/2$ is heavily suppressed by the removal of populations from these levels during state preparation. 
Excitation by $\sigma^{\pm}$ light is suppressed by the use of high quality polarisers in the clock laser beam and proper alignment with respect to the magnetic bias field.
We estimate line pulling effects on the clock transition to be below $1 \times 10^{-19}$.

The linear Zeeman shift is cancelled by averaging the frequencies of the two $|^3\mathrm{P}_0, m_F=\pm\frac{9}{2}\rangle \rightarrow |^1\mathrm{S}_0, m_F=\pm\frac{9}{2}\rangle$ clock transitions. 
However, a slowly fluctuating Zeeman splitting between these two transitions can result in a lock error. 
With a maximum observed drift per day of $\Delta \dot{\nu}_{\pm9/2} = 0.2 \, \mathrm{Hz/d}$ and a typical cycle time $t_\mathrm{seq} = 1.6 \, \mathrm{s}$, we estimate a maximum lock offset of ${\Delta \dot{\nu}_{\pm9/2} \cdot t_\mathrm{seq} / 2 \nu_0 = 4 \times 10^{-21}}$ for our interrogation sequence, which is negligible.

\subsection{Total systematic uncertainty and height reference}

\begin{table}[t]
    \begin{center}\small
	\begin{tabular}{p{4.5cm}  D{.}{.}{2}  D{.}{.}{2}}
		\hline
		\textbf{Frequency shift} & \multicolumn{1}{c}{\textbf{Value ($10^{-18}$)}} & \multicolumn{1}{c}{\textbf{Unc. ($10^{-18}$) }}\\
		\hline
		BBR from $T_\mathrm{shield}$		& -560.2 & 0.3 \\
		BBR from $T_\mathrm{out}$		& -5.6 & 0.3 \\
		Density						& +0.1 & 0.4 \\
		Lattice light					& +7.4 & 1.7 \\
		Second-order Zeeman			& -170.3 & 0.3 \\
		Background gas collision			& -2.3 & 1.0 \\
		Clock light					& -0.04 & 0.04 \\
		DC Stark						& -0.1 & 0.1 \\
		Servo error					& 0.0 & 0.2 \\
		Minor shifts					& 0.0 & < 0.1 \\
		\hline 
		\textbf{Total}					& -730.9 & 2.1 \\
		\hline
	\end{tabular}
    \end{center}
	\caption{Values and uncertainties for various frequency shifts in the transportable clock, for ${T_\mathrm{shield} = -100.18(2) \, ^{\circ}\mathrm{C}}$.}
	\label{tab:unc}
\end{table}

Values and uncertainties of the systematic shifts of the transportable clock are given in table~\ref{tab:unc}. 
The total estimated systematic uncertainty of the clock is $2.1 \times 10^{-18}$.

Applications that depend on the uncertainty budget such as clock comparisons also require correcting for differences in relativistic redshift.
This can be achieved by measuring the vertical height differences of the atomic samples with respect to external height reference markers.
The differential relativistic redshift is then determined using the local gravity acceleration, which is required to connect each clock's geopotential value to that of the respective reference marker, and the geopotential difference between markers.
This effect is usually treated separately from the clocks' other systematic effects.
In our case, the vertical height difference with respect to a reference marker next to the physics package is known with an uncertainty of $1 \, \mathrm{mm}$, which leads to a negligible additional fractional uncertainty contribution of about $1 \times 10^{-19}$.

\section{Absolute frequency measurement}\label{sec:AbsoluteFrequency}

The frequency of the transportable clock Sr4 was determined by comparison to the primary caesium fountain clocks CSF1 and CSF2 at PTB \cite{wey18} as described in \cite{sch20d}: 
The frequency of Sr4 is measured on shorter, interrupted intervals relative to a continuously running maser that serves as flywheel.
The average maser frequency on a longer, continuous interval is measured by the fountain clocks. 
Combining both measurements and correcting for the differential gravitational redshift between the clocks provides the absolute frequencies reported in table~\ref{tab:freq}. 
The centres of gravity of each clock's measurement versus the maser are adjusted to be the same such that the result is insensitive to a linear drift of the maser frequency in time.
Due to the noise of the flywheel oscillator's frequency, the average frequency measured by the strontium and caesium clocks may differ. 
The related extrapolation uncertainty $u_{\rm ext}$ is estimated using a sensitivity function based on the measurement intervals and a noise model of the flywheel \cite{gre16, sch20d}.
We have verified the previously observed noise assumptions on the hydrogen maser by comparisons with our optical clocks and find as the only deviation a reduced frequency flicker floor of $1 \times 10^{-16}$ (see \ref{app:maser} for details).
We treat the systematic corrections of each clock as correlated between measurements but not between different clocks, the extrapolation errors as correlated between measurements using the same time interval, and all other effects as uncorrelated.
Average results for each fountain clock and overall have been computed with weights that have been optimised by a least-squares algorithm to yield the lowest uncertainty including correlations (see \ref{app:correlations} for details).

We find average frequencies of the $^1\mathrm{S}_0 \rightarrow {^3\mathrm{P}_0}$ clock transition in $^{87}$Sr of $429 \, 228 \, 004 \, 229 \, 872.801(201)~\mathrm{Hz}$ using CSF1 and $429 \, 228 \, 004 \, 229 \, 872.975(86)~\mathrm{Hz}$ using CSF2, which are mostly limited by the systematic uncertainties of fountain clocks.
The overall average frequency $429 \, 228 \, 004 \, 229 \, 872.951(80)~\mathrm{Hz}$ has a fractional uncertainty of $1.9 \times 10^{-16}$ and is in agreement with previous measurements \cite{mar24a}.
Relevant correlation coefficients are reported in \ref{app:correlations}.

\begin{table*}
    \label{tab:freq}
    \makebox[\textwidth][c]{
    \scriptsize
	\begin{tabular}{l||rrl||lll|ll||lll|ll}
					\hline\hline
                    \multicolumn{4}{c||}{}			& \multicolumn{5}{c||}{CSF1}	& \multicolumn{5}{c}{CSF2} \\ \hline
			MJD		& $T_{\rm Sr}$& $u_{b,{\rm Sr}}$& $u_{\rm ext}$& $T_{\rm Cs}$& $u_{a,{\rm Cs}}$& $u_{b,{\rm Cs}}$ &  $\Delta\nu$ & $u$  &$T_{\rm Cs}$& $u_{a,{\rm Cs}}$& $u_{b,{\rm Cs}}$ & $\Delta\nu$  &  $u$ \\
	 		&	(s)					& ($10^{-18}$) & ($10^{-16}$) & (days)				 & 	\multicolumn{2}{c|}{($10^{-16}$)} & (Hz)& (Hz)&(days)				 & 	\multicolumn{2}{c|}{($10^{-16}$)}        & (Hz)& (Hz)\\ \hline	
60055	&	27604	&	17.0	&	1.6	&	2.52	&	8.2	&	2.8	&	872.79	&	0.38	&	2.52	&	3.3	&	1.7	&	873.11	&	0.17	\\
60060	&	108898	&	15.0	&	0.8	&	3.82	&	6.7	&	3.2	&	872.80	&	0.32	&	3.82	&	2.7	&	1.7	&	872.95	&	0.14	\\
60368	&	7679	&	3.5	&	3.0	&	2.09	&	9.3	&	4.7	&	872.36	&	0.47	&	2.09	&	3.8	&	1.7	&	872.75	&	0.22	\\
60371	&	78470	&	4.2	&	1.0	&	3.07	&	7.8	&	5.6	&	873.00	&	0.41	&	3.07	&	3.2	&	1.7	&	873.05	&	0.16	\\
60374	&	79960	&	5.1	&	0.9	&	3.63	&	7.1	&	5.4	&	872.99	&	0.39	&	3.63	&	3.3	&	1.7	&	873.04	&	0.16	\\
60385	&	69140	&	5.2	&	1.1	&	2.86	&	8.1	&	5.0	&	872.80	&	0.41	&	2.86	&	4.5	&	1.7	&	872.71	&	0.21	\\
60580	&	66629	&	8.5	&	1.1	&		&		&		&		&		&	3.16	&	3.1	&	1.7	&	873.06	&	0.16	\\
60592	&	157385	&	7.7	&	0.6	&		&		&		&		&		&	4.37	&	2.7	&	1.7	&	873.05	&	0.14	\\
60720	&	129451	&	2.4	&	0.8	&	3.50	&	7.3	&	1.9	&	872.82	&	0.33	&	3.50	&	3.0	&	1.7	&	872.83	&	0.15	\\
60725	&	225037	&	2.4	&	0.5	&		&		&		&		&		&	4.33	&	3.5	&	1.7	&	872.98	&	0.17	\\ \hline
																									
\textbf{Average}	&		&		&		&		&		&		&	872.801	&	0.201	&		&		&		&	872.975	&	0.086	\\ \hline\hline
	\end{tabular}
    }
    \caption{\label{tab:data}Summary of the relevant information on the absolute frequency measurements for the respective measurement intervals labelled by the Modified Julian Date (MJD, centre of gravity). $\Delta\nu$ is the value to be added to $429\,228\,004\,229\,000$~Hz to get the frequency $\nu_{\rm Sr}$ of the $5\mathrm{s}^2 \, ^1\mathrm{S}_0 \rightarrow 5\mathrm{s}5\mathrm{p} \, ^3\mathrm{P}_0$ transition in $^{87}$Sr measured by the fountain clocks CSF1 and CSF2, respectively. The statistical uncertainty contribution from the Sr frequency standard is below $7 \times 10^{-18}$ for each measurement and therefore negligible.}	
\end{table*}
\normalsize

\section{Summary and outlook}\label{sec:Outlook}

We have described a transportable strontium lattice clock with a BBR shift uncertainty of $4.0 \times 10^{-19}$, getting the BBR shift under better control compared to state-of-the-art lab-based strontium lattice clocks, where it often represents the largest systematic uncertainty contribution. This was achieved by interrogating the atoms in a thermal shield with low homogeneous temperature and carefully characterised apertures to the outside. The low self-comparison instability of $7 \times 10^{-16}/\sqrt{\tau/\mathrm{s}}$ using the transportable clock laser \cite{her22} allows quick recharacterizations of systematic effects. This enables a total systematic uncertainty of $2.1 \times 10^{-18}$, both on- and off-campus, that is comparable to the current relativistic redshift uncertainty at the geodetically best-surveyed locations, using state-of-the-art geodetic techniques \cite{den17}, or even below that \cite{pav17, gro18a, lee24}. Furthermore, by comparing the clock to the caesium fountains CSF1 and CSF2 at PTB, we have measured the frequency of the $^1\mathrm{S}_0 \rightarrow {^3\mathrm{P}_0}$ clock transition of $^{87}$Sr.

\cite{icon24} presents the first measurement campaign of this clock. Throughout following measurement campaigns, the uncertainty as well as the stability of the clock could be improved to reach the clock performance presented here. It is planned to participate in inter-institute frequency comparisons at the $10^{-18}$ level. This transportable clock is therefore expected to facilitate the development of cm-level chronometric geodesy \cite{meh18, tan21b, tak22}. A further improvement of the clock's uncertainty and reproducibility appears to be feasible by extended investigations of the lattice light shift \cite{kim23} and the density shift \cite{aep22}.

\ack
We thank Richard Hobson for help in designing the pyramid MOT. Regarding the construction of the pyramid MOT mirrors, we thank Andr\'e Uhde and Mandy Rindermann for manufacturing the copper substrates, Rudolf Mee\ss~and Stefan Verh\"ulsdonk for the surface processing of their optical surfaces, and Daniel Hagedorn and Steffen Wei\ss~for applying their coatings. We also express our gratitude to Andr\'e Uhde for machining the BBR shield, and to Michael Neugebauer, Daniel Bennett, Erik Jansson and Elena Jordan for helping in the characterization of its hole radii. We thank Kilian Stahl and Joshua Klose for operating the laboratory clock Sr3 as a stable reference, and Uwe Sterr and Thomas Legero for providing it with ultrastable laser light.

We acknowledge support by the Deutsche Forschungsgemeinschaft (DFG, German Research Foundation) under Germany’s Excellence Strategy -- EXC-2123 QuantumFrontiers -- Project-ID 3908379.67 and SFB 1464 TerraQ -- Project-ID 434617780 -- within project A04. This work was partially supported by the Max Planck–RIKEN–PTB Center for Time, Constants and Fundamental Symmetries. This work has received funding from the European Partnership on Metrology, co-financed by the European Union’s Horizon Europe Research and Innovation Programme and by the Participating States, under grant number 22IEM01 TOCK.

\section*{Author Contributions}
I.N., C.V., T.L., and C.L. designed the transportable clock including the BBR shield. I.N., C.V., and T.L. built, operated and characterised the transportable clock. J.R., N.P., and S.W. operated the caesium fountain clocks, and E.B. operated a frequency comb connecting them to the transportable clock. C.L., S.D., and S.W. analysed the absolute frequency measurement results. I.N., C.V., T.L., S.D., and C.L. wrote the manuscript. All authors discussed the results.

\appendix
\setcounter{table}{0}
\renewcommand{\thetable}{A\Roman{table}}
\setcounter{figure}{0}
\renewcommand{\thefigure}{A\arabic{figure}}
\renewcommand{\theequation}{A\arabic{equation}}

\section{Level scheme}\label{app:LevelScheme}

A level scheme for the addressed electronic transitions described in this manuscript is shown in figure \ref{fig:level_scheme}.

{\begin{figure*}[h]
\centering
\includegraphics[width=0.7\linewidth]{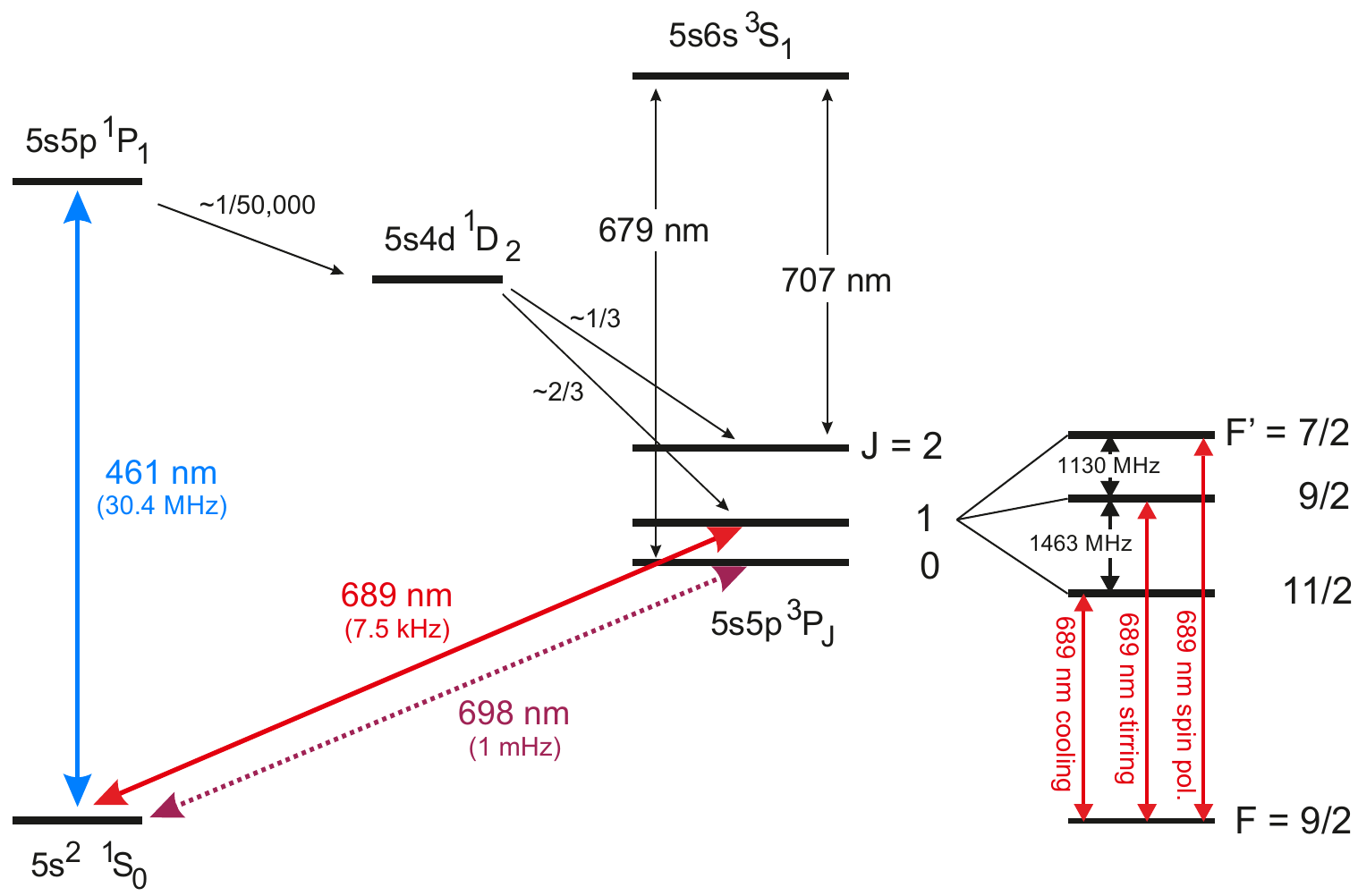}
\caption{Partial electronic level scheme of $^{87}$Sr. Linewidths for electronic transitions from the ground state are calculated from the upper state lifetimes for the $^1\mathrm{P}_1$ \cite{hei20}, $^3\mathrm{P}_1$ \cite{aep24_sup} and $^3\mathrm{P}_0$ \cite{mun21} states. Additionally shown are partial decay rates to \cite{hun86} and from \cite{bau85a} the $^1\mathrm{D}_2$ state, and the hyperfine splittings of the $^3\mathrm{P}_1$ state \cite{put63}.}
\label{fig:level_scheme}
\end{figure*}}

\section{BBR shield bore radii}\label{app:SolidAngle}

\setcounter{table}{0}
\renewcommand{\thetable}{B\Roman{table}}
\setcounter{figure}{0}
\renewcommand{\thefigure}{B\arabic{figure}}
\renewcommand{\theequation}{B\arabic{equation}}

In the estimation of the influence of external BBR on atoms in the shield, the determination of the solid angle $\Omega$ under which the atoms see the BBR shield holes, is critical.
Here the radii of the bores are the parameters the most difficult to determine.
The radii of the holes 
were measured using a coordinate measuring machine (CMM) with a ruby ball stylus with diameter $300 \, \mu\mathrm{m}$ before the inner surfaces were coated.
The positions of 16 points on each of three planes at depths of \mbox{100 $\mu$m}, \mbox{500 $\mu$m} and \mbox{900 $\mu$m} have been measured for each hole, from which the radius at each depth was determined.
It was found that the hole radii shrink by about \mbox{8 $\mu$m} towards the inside.
As the radii at the outer hole edges are relevant for determining the frequency shift of the atoms due to room temperature BBR, we linearly extrapolated to the outer edge.
We find radii $r_{\mathrm{CMM},1}^\mathrm{out} = 0.520 \, 5(15) \, \mathrm{mm}$ and $r_{\mathrm{CMM},2}^\mathrm{out} = 0.522 \, 5(15) \, \mathrm{mm}$.

Additionally, we measured the radii along eight orientations for each hole by a measuring microscope (MM) before and after applying the black coating on the inner hole surfaces and cold-welding the shield lid onto the main body of the shield, $r_{\mathrm{MM},i}^\mathrm{before}(\alpha)$ and $r_{\mathrm{MM},i}^\mathrm{after}(\alpha)$.
Due to the coating and an emerging slight ellipticity of the holes due to the pressing of the top lid on the main body, we observed a reduction of radii with respect to the average bore radii determined with the measurement microscope before.
While using the measurement microscope has the advantage of enabling non-tactile measurements which do not damage the coating, it provides less accurate results for deep holes. 
This data are corrected by the difference between the coordinate measuring machine results and the initial average measurement microscope results, $\Delta r_{\mathrm{CMM},i} = r_{\mathrm{CMM},i}^\mathrm{out} - \overline{r_{\mathrm{MM},i}^\mathrm{before}}$, which corresponds to $+3.5 \, \mu\mathrm{m}$ for hole 1 and $+6.5 \, \mu\mathrm{m}$ for hole 2.

{\begin{figure}[tb]
\centering
\includegraphics[width=0.5\linewidth]{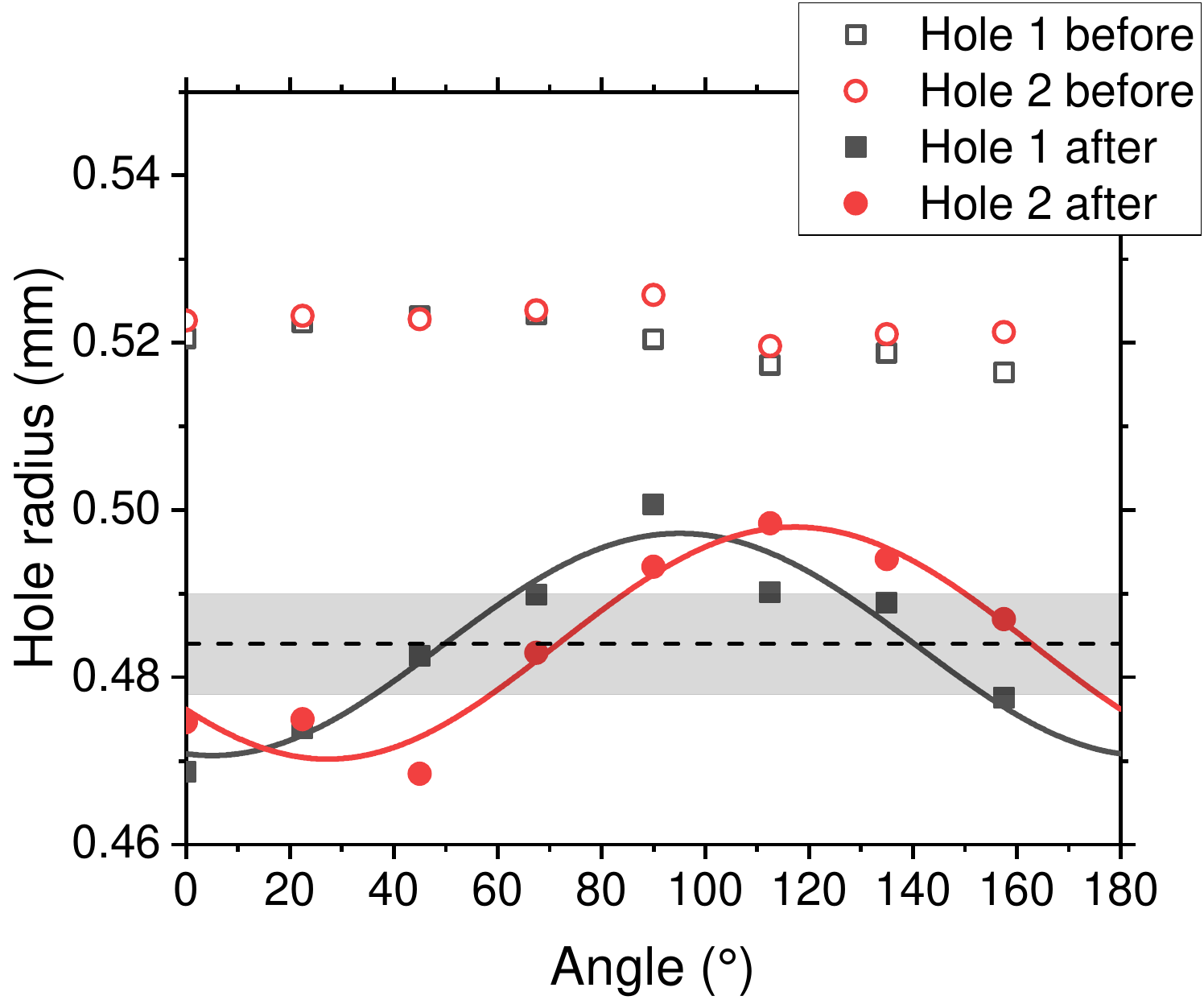}
\caption{The radii of the two holes of the BBR shield plotted over the angle between the axis towards the BBR shield top lid and the direction of measurement. They are shown both before and after the black coating and the connection of the two BBR shield parts. The latter are fitted with the formula for the angle-dependent radius of an ellipse \mbox{$r(\alpha) = \left( \left[ r_\mathrm{min} \cos \left( \pi \frac{\alpha - \alpha_0}{180^{\circ}} \right) \right]^2 +  \left[ r_\mathrm{max} \sin \left( \pi \frac{\alpha - \alpha_0}{180^{\circ}} \right) \right]^2 \right)^{1/2}$}.
The solid lines show the respective fit results, and the dashed line the effective radius $r_{\mathrm{eff}}$ with its uncertainty (grey area).}
\label{fig:hole_radii}
\end{figure}}

The resulting data are shown in figure~\ref{fig:hole_radii}. 
The effective radius for the holes after coating and connection is calculated by $r_{\mathrm{eff},i} = \left( r_{\mathrm{min},i} \, r_{\mathrm{max},i} \right)^{1/2}$, with the fitted minimum and maximum radii $r_{\mathrm{min},i}$ and $r_{\mathrm{max},i}$. 
We find for both bores $r_\mathrm{eff} = 0.484(6) \, \mathrm{mm}$, with the uncertainty mostly stemming from the measurements with the measurement microscope.

The uncertainty of the offset of the atoms from the centre of the BBR shield is estimated to be \mbox{0.5 mm}. 
This is based on the measurement of the BBR shield hole positions shown in figure~\ref{fig:InterleavedResults}~(a).
The uncertainty of the BBR shield length is estimated to be \mbox{0.1 mm}.
Close to the centre of the BBR shield, where $|z|,r_1,r_2 \ll \frac{\ell}{2}$, (\ref{eq:SolidAngle}) simplifies to
\begin{equation}\label{eq:SolidAngleSimple}
\frac{\Omega}{4\pi} \approx \frac{r_1^2}{4 \left( z + \frac{\ell}{2} \right)^2} + \frac{r_2^2}{4 \left( z - \frac{\ell}{2} \right)^2}~,
\end{equation}
from which we calculate the uncertainty contributions in table~\ref{tab:SolidAngleUnc}.

\begin{table*}[t]
    \begin{center}\small
	\begin{tabular}{c | c | c | c}
		\hline
		\textbf{Quantity} & \textbf{Value} & \textbf{Uncertainty} & \textbf{$\delta(\Omega/4\pi)$}\\
		\hline
		Hole radii $r_1$	 and $r_2$	& $484 \, \mu\mathrm{m}$ (both) & $6 \, \mu\mathrm{m}$ (both) &  $2.5 \times 10^{-5}$ \\
		Atom position $z$	& $0 \, \mathrm{mm}$ & $0.5 \, \mathrm{mm}$ & $0.9 \times 10^{-5}$ \\
		BBR shield length $\ell$		& $20 \, \mathrm{mm}$ & \mbox{0.1 mm} & $1.2 \times 10^{-5}$ \\
		\hline 
		\multicolumn{3}{r}{\textbf{Total:}}		& \textbf{$2.9 \times 10^{-5}$} \\
		\hline
	\end{tabular}
    \end{center}
	\caption{Uncertainty contributions of the fractional solid angle of the two BBR shield holes.}
	\label{tab:SolidAngleUnc}
\end{table*}

We note that the radii shrink by up to 0.3\% if the copper shield is cooled from room temperature to cryogenic temperatures \cite{bee55}, i.e. by up to $1.5~\mu\mathrm{m}$ -- however, as also the length of the BBR shield shrinks in the same way, the $\Omega$ stays constant to a good approximation.

\section{Maser noise model}\label{app:maser}

\setcounter{table}{0}
\renewcommand{\thetable}{C\Roman{table}}
\setcounter{figure}{0}
\renewcommand{\thefigure}{C\arabic{figure}}
\renewcommand{\theequation}{C\arabic{equation}}

{\begin{figure}[h]
\centering
\includegraphics[width=0.65\linewidth]{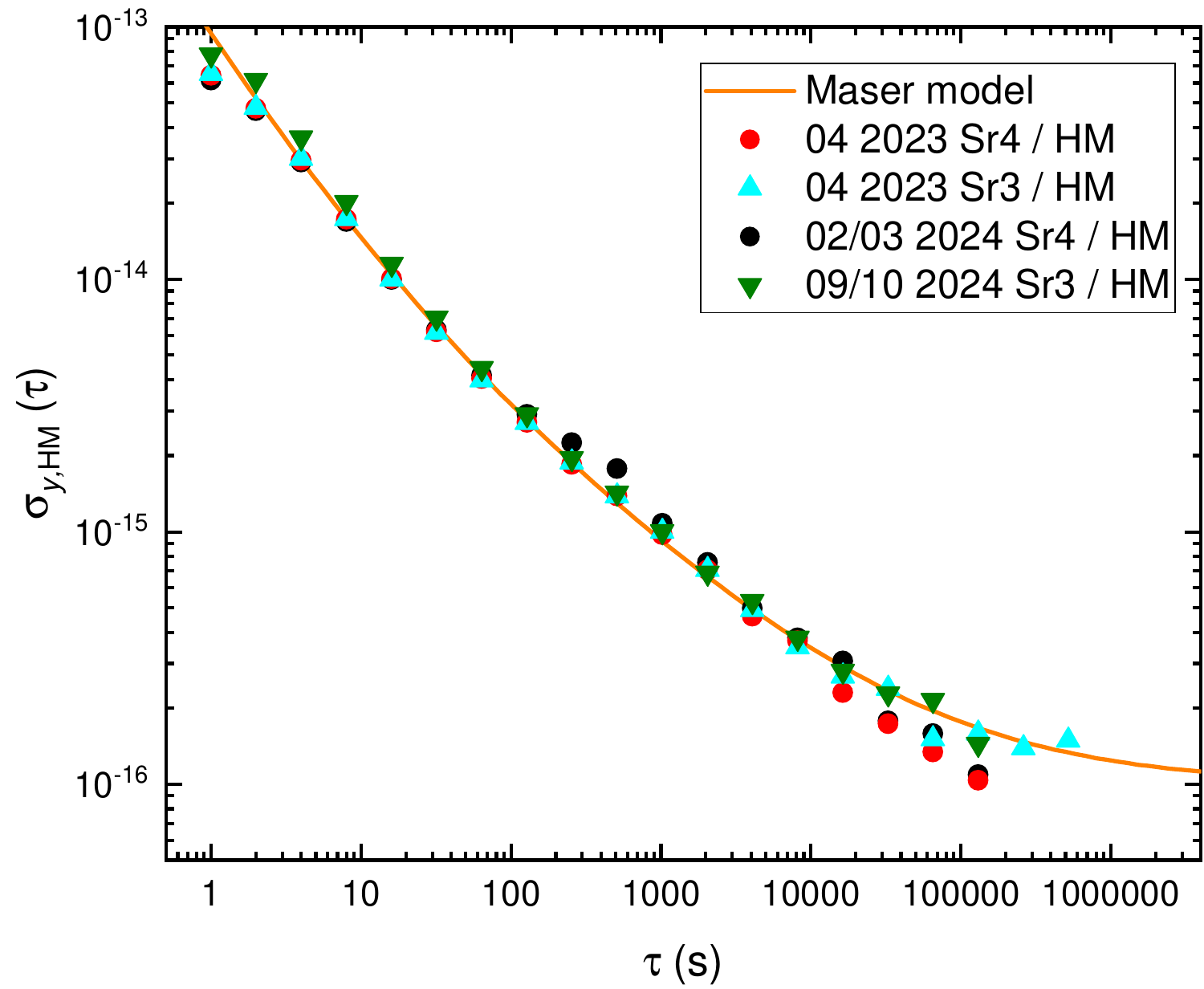}
\caption{Allan deviations showing the hydrogen maser's frequency stability as measured by Sr lattice clocks at PTB. The curve depicts the maser instability derived from the maser noise model.}
\label{fig:ADEV_HM}
\end{figure}}

For the measurement intervals listed in table~\ref{tab:freq}, we have modelled the noise spectrum of the hydrogen maser that was used as flywheel oscillator by the power law expansion $S_{y} = \sum_{\alpha=-1}^1{h_\alpha\,f^\alpha}$ of the single-sided power spectral density $S_y(f)$ by adding flicker frequency (FFN), white frequency (WFN) and flicker phase noise (FPN) contributions. The coefficients $h_\alpha$ are adjusted such that the calculated ADEV $\sigma_{y, \rm{HM}}$ \cite{daw07,ben15} matches the observations in figure~\ref{fig:ADEV_HM}. We find the coefficients 
\begin{eqnarray}
	h_1 = 4.3 \times 10^{-26}\;\textrm{Hz}^{-2} &\Rightarrow& \; \sigma_{y,\rm{FPN}} = 7 \times 10^{-14} \; \mathrm{s}/\tau \nonumber \\
    h_0 = 1.2 \times 10^{-27}\;\textrm{Hz}^{-1} &\Rightarrow& \; \sigma_{y,\rm{WFN}} = 2.4 \times 10^{-14} \; \sqrt{\mathrm{s}/\tau} \nonumber \\
    h_{-1} = 7.2 \times 10^{-33} &\Rightarrow& \; \sigma_{y,\rm{FFN}} = 1 \times 10^{-16} \label{eq:HM}		
\end{eqnarray}
using a cut-off frequency of 0.5~Hz.

\section{Correlation analysis and averaging of absolute frequency measurements}\label{app:correlations}

\setcounter{table}{0}
\renewcommand{\thetable}{D\Roman{table}}
\setcounter{figure}{0}
\renewcommand{\thefigure}{D\arabic{figure}}
\renewcommand{\theequation}{D\arabic{equation}}

The evaluation of the weighted average values reported for the absolute frequency in section~\ref{sec:AbsoluteFrequency} (see also table~\ref{tab:freq}) takes into account the correlations between the measurements.
We follow the general notation introduced in \cite{mar20}.
To facilitate comparisons to other results, we analyse the correlations of the measurements to specific error contributions, $\Delta_k$, i.e., the random variables underlying uncertainty contributions $u_k$, such as the systematic uncertainties of the clocks.
These error contributions are -- exactly or approximately -- either fully correlated or uncorrelated between measurements:
the systematic errors of each clock are treated as fully correlated between measurements but uncorrelated between clocks; extrapolation errors are fully correlated between all measurements within the same interval and treated as uncorrelated otherwise; the fountain clocks' statistical errors are uncorrelated; we neglect the statistical error of the optical clock, which can be considered part of the extrapolation error.
The covariance $u(\nu_i, \Delta_k)$ between a measurement $\nu_i$ and a relative error $\Delta_k$ is then given by the respective uncertainty contribution listed in table~\ref{tab:freq} or zero.
We note that the covariances are negative for the systematic errors of the fountain clocks as the frequency measured for the optical clock decreases if their frequency is increased.

Given a set of weights $w_i$ where $\sum_i w_i = 1$, the weighted average value and its covariances with respect to the independent quantities $\Delta_k$ are then given by:
\begin{eqnarray}
    \bar{\nu}                                          & = & \sum_{i=1}^N w_i \nu_i \\
    u(\bar{\nu}, \Delta_k)                             & = & \sum_{i=1}^N w_i u(\nu_i, \Delta_k)
\end{eqnarray}
The uncertainty is
\begin{equation}
    u(\bar{\nu}) = \sqrt{\sum_{k=1}^K u(\bar{\nu}, \Delta_k)^2}
    \label{eq:average-uncertainty}
\end{equation}
since the $\Delta_k$ account for all relevant independent quantities.
Finally, the correlation coefficients between two quantities, $q_i$ and $q_j$, can readily be derived using \cite{mar20}
\begin{equation}
    r(q_i, q_j) = \frac{u(q_i, q_j)}{u(q_i) u(q_j)}\mbox{,}
\end{equation}
where the uncertainty of the relative errors $u(\Delta_k)$ is one.

We determine optimised sets of weights $w_i$ for the full data set as well as the subsets measured against a specific fountain by minimising the uncertainty $u(\bar{\nu})$ using a least-squares algorithm with equal weights as initial estimate.
The weights are summarised in table~\ref{tab:abs-freq-weights}.
For comparison, we also include the weights that would result from simply using the combined statistical and extrapolation variances ($w_i^\prime \propto 1 / (u_\mathrm{a}^2 + u_\mathrm{ext}^2)$) for the CSF1 and CSF2 subsets.
The optimised weights reproduce the fixed weights well for the CSF2 subset, which is expected because the systematic uncertainty of the fountain is nearly constant across the measurements.
The other fountain clock's systematic uncertainty by contrast varies substantially between measurements.
The optimised set thus assigns higher weights to the more accurate measurements in this case in comparison to the fixed weights, which leads to a moderate reduction of the uncertainty (by about $6\,\%$).

Concerning the uncertainty of the overall average, we note that both the increase by including the correlations between the subset averages that arise from extrapolation errors ($<1\,\%$) as compared to the weighted average used in \cite{sch20d} and the reduction by adopting the optimised weights (about $1\,\%$) are marginal.
We report the relevant correlation coefficients in table~\ref{tab:abs-freq-correlations}.

\begin{table}[ht]
    \begin{center}\scriptsize
	\begin{tabular}{l|r|r|r|r|rr|rr}
		\hline
		\multicolumn{1}{c|}{\bfseries MJD}    & \multicolumn{8}{c}{\bfseries Weight}                                                                                                                                                                                                                                                                                      \\
		\multicolumn{1}{c|}{} & \multicolumn{2}{c|}{\bfseries CSF1 average}                                    & \multicolumn{2}{c|}{\bfseries CSF2 average}                                    & \multicolumn{4}{c}{\bfseries CSF1+CSF2 average}                                                                                                            \\
		                                      & \multicolumn{1}{c|}{\bfseries simple} & \multicolumn{1}{c|}{\bfseries optimised} & \multicolumn{1}{c|}{\bfseries simple} & \multicolumn{1}{c|}{\bfseries optimised} & \multicolumn{2}{c|}{\bfseries simple}                                        & \multicolumn{2}{c}{\bfseries optimised}                                       \\
		                                      & \multicolumn{1}{c|}{\bfseries (CSF1)} & \multicolumn{1}{c|}{\bfseries (CSF1)}  & \multicolumn{1}{c|}{\bfseries (CSF2)} & \multicolumn{1}{c|}{\bfseries (CSF2)}  & \multicolumn{1}{c}{\bfseries (CSF1)} & \multicolumn{1}{c|}{\bfseries (CSF2)} & \multicolumn{1}{c}{\bfseries (CSF1)} & \multicolumn{1}{c}{\bfseries (CSF2)} \\
		\hline
		60055 & 0.123 & 0.181 & 0.085 & 0.084 & 0.015 & 0.075 & 0.026 & 0.068\\
		60060 & 0.189 & 0.248 & 0.145 & 0.144 & 0.022 & 0.128 & 0.038 & 0.121\\
		60368 & 0.090 & 0.067 & 0.049 & 0.049 & 0.011 & 0.043 & 0.008 & 0.039\\
		60371 & 0.140 & 0.055 & 0.102 & 0.102 & 0.016 & 0.090 & 0.009 & 0.088\\
		60374 & 0.168 & 0.079 & 0.097 & 0.098 & 0.020 & 0.086 & 0.013 & 0.083\\
		60385 & 0.129 & 0.080 & 0.054 & 0.054 & 0.015 & 0.047 & 0.013 & 0.046\\
		60580 &       &       & 0.107 & 0.107 &       & 0.095 &       & 0.093\\
		60592 &       &       & 0.149 & 0.149 &       & 0.132 &       & 0.129\\
		60720 & 0.160 & 0.290 & 0.120 & 0.120 & 0.019 & 0.106 & 0.045 & 0.101\\
		60725 &       &       & 0.092 & 0.092 &       & 0.081 &       & 0.080\\
		\hline
	\end{tabular}
    \end{center}
    \caption{
        Weights of measurement results in the absolute frequency averages for the full data set and for the CSF1 and CSF2 subsets.
        In addition to the optimised weights, the weights resulting from simple weighting strategies are given for comparison; see the main text for details.
    }
    \label{tab:abs-freq-weights}
\end{table}

\begin{table}[ht]
    \begin{center}\small
	\begin{tabular}{l|l|D{.}{.}{2}}
		\hline
		$q_i$ & $q_j$ & \multicolumn{1}{l}{$r(q_i, q_j)$} \\
		\hline
		$\bar{\nu} (\mathrm{Sr}|\mathrm{CSF1})$ & $\Delta_{\mathrm{b}, \mathrm{Sr}}  $    &  0.019 \\
		$\bar{\nu} (\mathrm{Sr}|\mathrm{CSF1})$ & $\Delta_{\mathrm{b}, \mathrm{CSF1}}$    & -0.706 \\
		$\bar{\nu} (\mathrm{Sr}|\mathrm{CSF2})$ & $\Delta_{\mathrm{b}, \mathrm{Sr}}  $    &  0.038 \\
		$\bar{\nu} (\mathrm{Sr}|\mathrm{CSF2})$ & $\Delta_{\mathrm{b}, \mathrm{CSF2}}$    & -0.845 \\
		$\bar{\nu} (\mathrm{Sr})$               & $\Delta_{\mathrm{b}, \mathrm{Sr}}  $    &  0.041 \\
		$\bar{\nu} (\mathrm{Sr})$               & $\Delta_{\mathrm{b}, \mathrm{CSF1}}$    & -0.271 \\
		$\bar{\nu} (\mathrm{Sr})$               & $\Delta_{\mathrm{b}, \mathrm{CSF2}}$    & -0.776 \\
		$\bar{\nu} (\mathrm{Sr})$               & $\bar{\nu} (\mathrm{Sr}|\mathrm{CSF1})$ &  0.397 \\
		$\bar{\nu} (\mathrm{Sr})$               & $\bar{\nu} (\mathrm{Sr}|\mathrm{CSF2})$ &  0.923 \\
		\hline
	\end{tabular}
    \end{center}
	\caption{Correlation coefficients $r$ of the average transition frequencies of the full data set, $\bar{\nu}(\mathrm{Sr})$, and the subsets using the same fountain clock, $\bar{\nu} (\mathrm{Sr}|i)$, with respect to each other and to the atomic clocks' systematic relative errors, $\Delta_{\mathrm{b},j}$.
    }
	\label{tab:abs-freq-correlations}
\end{table}

\section*{References}
\bibliographystyle{iopart-num}
\bibliography{texbi431}

\end{document}